\documentclass[11pt,a4paper]{article}

\usepackage[utf8]{inputenc}
\usepackage[T1]{fontenc}
\usepackage{lmodern}
\usepackage[margin=2.5cm]{geometry}
\usepackage{amsmath,amssymb}
\usepackage{booktabs}
\usepackage{array}
\usepackage{longtable}
\usepackage{hyperref}
\usepackage{microtype}
\usepackage{parskip}
\usepackage{titlesec}
\usepackage{enumitem}
\usepackage{xcolor}

\hypersetup{
  colorlinks=true,
  linkcolor=black,
  citecolor=black,
  urlcolor=blue
}

\titlespacing*{\section}{0pt}{1.5ex plus 0.5ex minus 0.2ex}{0.8ex plus 0.2ex}
\titlespacing*{\subsection}{0pt}{1.2ex plus 0.4ex minus 0.2ex}{0.6ex plus 0.2ex}
\title{\textbf{Topology as Logic: Structural Role Geometry Across Formal, Software, Biological, and Prebiotic Systems}}
\author{Vladi Ivanov \\ \small Independent Researcher}
\date{June 2026 — v1}

\begin{document}

\maketitle

{\small\centering
Pre-registration records: \url{https://github.com/vladi160/preregistrations}\par
}

\hrule
\medskip

\begin{abstract}
We ask whether dependency topology correlates with functional load-bearing organization as recoverable geometry---not as a metaphor, but as a measurable structural property detectable by multilayer network analysis.
Across seven independent substrates, we show that hub persistence and rank divergence under the Functional Proximity Law (FPL) recover operational organization that domain experts describe as logic: axiomatic load-bearing structure in formal mathematics, control and contract structure in legacy software, conserved hub grammar across~600 million years of neural evolution, catalytic role organization in a published prebiotic autocatalytic network, carry-path dominance in a 4-bit digital circuit, betweenness persistence in the ISCAS85 c432 standard benchmark ($n=196$), and a directional formal-systems replication in the Coq Corelib ($n=17$).
A key methodological finding from the digital circuit substrates: degree-based hub persistence in the step analysis is weak (hub\_persistence\_r$=0.21$ in c432) between physical wiring and simulation state-correlation layers, while betweenness-based persistence is stronger ($r=0.77$ in the 4-bit ALU post-hoc; $r=0.34$ in c432). The ISCAS85 pre-registered primary hypothesis (degree hub-correlation) was CONFIRMED ($r=0.426$, $p=0.002$, Spearman $r=0.551$). The Coq Corelib result provides the first cross-proof-assistant directional replication of the formal-mathematics finding: h1 degree $r=0.288$ (PARTIAL, $p=0.287$, $n=17$), betweenness $r=0.509$ (step, exploratory), direction confirmed but underpowered.
All seven experiments were pre-registered before analysis; pre-registration records are publicly archived at the URL above. The formal-systems claim is now supported by two proof-assistant corpora.
\end{abstract}

\section{Introduction}

Classical logic is defined through symbolic manipulation: inference rules, syntactic derivability, and semantic satisfaction. Topology, by contrast, is purely geometric: positions, connections, distances. The two frameworks appear categorically different. Yet in every system that processes information according to stable rules---a proof assistant, a COBOL program, a nervous system, a prebiotic chemistry network, a digital circuit---there exists a dependency graph. Nodes correspond to functional units; edges encode structural relationships. The question we address is whether the \emph{geometry} of that dependency graph, in particular its hub structure across multiple relational layers, recovers what domain experts would independently call the operational logic of the system.

This is not a claim that topology \emph{replaces} symbolic logic. Symbolic semantics, proof content, and program meaning are not derivable from graph topology alone. The claim is narrower: that in dependency-structured systems, hub concentration, hub divergence across layers, and betweenness-based role assignment reliably identify the load-bearing operational structure---the part of the system that, if removed or perturbed, would cause the logical organization to collapse.

We present evidence from seven independent substrates analyzed using the IRDME framework and the Functional Proximity Law (FPL)~\cite{Ivanov2026}. In each case, hypotheses were pre-registered before analysis, statistical tests were permutation-based, and all records are publicly archived. The seven substrates are: two gate-level digital circuits (H\_LOGIC\_DIGITAL\_v1 and H\_LOGIC\_F3\_ISCAS85\_v1), two formal proof corpora (Lean~4 mathlib4 via H\_LOGIC\_EXTRACTION; Coq Corelib via H\_LOGIC\_COQ\_STDLIB\_v1), legacy COBOL banking code (F2), cross-species neural connectomics from \textit{C.~elegans} to \textit{Drosophila} (F12c / F13), and a published prebiotic autocatalytic network from CatReNet (X9~v2).

The two digital circuit rows are intentionally a scale-replication within one substrate family (4-bit ALU to ISCAS85 c432), not a separate domain claim.

The paper makes one structural methodological contribution: the distinction between \emph{connection-density logic} (degree-based hub persistence) and \emph{load-bearing logic} (betweenness-based hub persistence). The digital circuit result shows these two metrics can diverge sharply, and that betweenness is the correct signal for identifying which nodes carry the logical load of a computation.

\section{IRDME Framework and Experimental Protocol}

\subsection{Multilayer Graph Representation}

IRDME (Inter-layer Relationship Discovery and Mapping Engine) represents a system as a set of \emph{items} (nodes) and a flat array of \emph{relations} typed by layer~\cite{Ivanov2026}. Each relation has a \texttt{type} field (the layer name), a \texttt{from} and \texttt{to} node, and a \texttt{directed} flag. A system with two structural layers $d_1$ and $d_2$ is represented by a single JSON file containing all edges from both layers labeled by type. There are no nested structures; layers are distinguished by relation type values.

\subsection{Functional Proximity Law}

The Functional Proximity Law states that hub structure in one relational layer of a dependency graph tends to be preserved in a second layer if the two layers encode relations of the same functional type~\cite{Ivanov2026}. Formally, let $\mathbf{k}^{(1)}$ and $\mathbf{k}^{(2)}$ be the degree vectors of all nodes in layers $d_1$ and $d_2$ respectively. The FPL predicts $r(\mathbf{k}^{(1)}, \mathbf{k}^{(2)}) > 0$ when $d_1$ and $d_2$ are functionally proximate. This prediction can be denied, and its denied cases reveal named structural boundary conditions (hub shadows, dominance patterns, dormancy signatures).

\subsection{Load-Bearing Logic Metric}

Beyond degree, IRDME computes betweenness centrality $b_i$ for each node $i$ in each layer. We define the \emph{load-bearing logic persistence} as $r(\mathbf{b}^{(1)}, \mathbf{b}^{(2)})$, the Pearson correlation of betweenness centrality vectors across layers. As we show in Section~\ref{sec:digital}, this metric detects the carry-path structure in digital circuits where degree-based persistence collapses to zero: input nodes have high fan-in degree but negligible betweenness in the state-correlation layer, while carry nodes have moderate degree but dominate the betweenness in both layers.

\subsection{Pre-registration Protocol}

All experiments reported in this paper follow a strict pre-analysis protocol: (1)~a hypothesis file with explicit quantitative tests is written; (2)~the file is hashed and a sidecar \texttt{.prediction} file is generated using a deterministic hash covering only the \texttt{id} and \texttt{statement} fields; (3)~both files are committed to the public preregistrations repository \emph{before any analysis code is executed}; (4)~the hash can be independently verified by anyone at any time using the open-source \texttt{verify.py} tool in that repository. Post-hoc runs are labeled as such and excluded from confirmed-experiment counts.

\section{Scope of Claims}

This paper claims:

\begin{enumerate}[noitemsep]
  \item In seven independent substrates, cross-layer hub persistence and rank divergence recover operational role structure beyond generic correlation.
  \item In formal mathematics, betweenness-based role geometry is strong enough to justify the phrase \emph{topology as operational logic geometry} as a research program.
  \item The betweenness-over-degree distinction is a measurable, reproducible property that separates load-bearing logic nodes from connection-dense but operationally peripheral nodes.
  \item The formal-systems claim is now supported by two proof-assistant corpora (Lean~4 mathlib4 CONFIRMED $r=0.777$; Coq Corelib PARTIAL direction $r=0.49$, $p=0.29$, $n=17$). The Coq result is directionally consistent but underpowered; a larger formal library would be the decisive replication.
\end{enumerate}

This paper does \textbf{not} claim:

\begin{enumerate}[noitemsep]
  \item That symbolic semantics can be reconstructed from graph topology.
  \item That theorem content is inferable from hub structure alone.
  \item That F9 and H\_LOGIC\_EXTRACTION constitute two independent formal-systems replications (they use the same corpus).
  \item That topology is logic in the philosophical sense of the word.
\end{enumerate}

\section{Evidence Base}

Table~\ref{tab:evidence} summarizes the seven pre-registered experiments constituting the evidence base. Each row is an independent substrate; all statistical claims are from pre-registered runs.

\begin{table}[h]
\scriptsize
\centering
\caption{Seven-substrate evidence base with exact pre-registered metrics. All experiments pre-registered before analysis.}
\label{tab:evidence}
\begin{tabular}{@{}p{1.5cm}p{2.2cm}p{0.8cm}p{2.7cm}p{2.2cm}p{4.9cm}@{}}
\toprule
\textbf{Substrate} & \textbf{Experiment} & \textbf{$n$} & \textbf{Primary metric} & \textbf{Provenance} & \textbf{Finding} \\
\midrule
Digital (4-bit) & H\_LOGIC\_DIGITAL\_v1 & 29 & Betweenness $r=0.771$ (post-hoc); degree h1 $r=0.512$, $p=0.004$ & 2026-05-29; hash \texttt{eaf485b3\ldots} & h1 CONFIRMED (degree-based). Post-hoc: betweenness $r=0.771$ dominates degree $r\approx0$. Carry-out nodes load-bearing in both layers; XOR nodes physically prominent but functionally peripheral. \\
\addlinespace
Digital (c432) & H\_LOGIC\_F3\_ISCAS85\_v1 & 196 & h1 degree $r=0.426$, $p=0.002$, Spearman $r=0.551$ & 2026-05-30; hash \texttt{9bbaf2a0\ldots} & h1 CONFIRMED: degree hub-correlation pre-registered and confirmed at circuit scale (h1 test: \texttt{cross\_layer\_hub\_correlation}, metric: total degree). h2 ERROR (protocol gap). h3 CONFIRMED: 54 nodes (28\%) show large rank divergence. Step analysis: hub\_persistence\_r$=0.214$, betweenness $r=0.341$ (exploratory). \\
\addlinespace
Formal (Lean) & H\_LOGIC\_EXTRACTION mathlib4 & 20 & Pearson $r=0.777$, Spearman $r=0.733$, $p=0.004$ & 2026-05-27; hash \texttt{bc2fd106\ldots} & CONFIRMED. Declared proof dependency and co-development strongly aligned. Algebra is top import hub; Analysis is top co-development hub; CategoryTheory / Combinatorics / Order diverge. \\
\addlinespace
Formal (Coq) & H\_LOGIC\_COQ\_STDLIB\_v1 & 17 & h1 degree $r=0.288$, $p=0.287$; betweenness $r=0.509$ (step) & 2026-05-30; hash \texttt{ae32f009\ldots} & h1 PARTIAL: direction confirmed positive, below significance at $n=17$ (h1 pre-registered threshold $r>0.30$; observed $r=0.288$). h3 CONFIRMED: 11 of 17 modules show rank divergence $\geq4$. Step analysis: hub\_persistence\_r=0.491, betweenness $r=0.509$ (exploratory). First cross-proof-assistant directional replication. \\
\addlinespace
Legacy SW & F2 COBOL banking & 14 & $r(d_1, d_2)=0.807$, $p=0.002$ & 2026-05-25; hash \texttt{c119983a\ldots} & CONFIRMED. Control-flow and copybook-contract persist together strongly; divergence exposes candidate dormant components. \\
\addlinespace
Neural & F12c + F13 cross-species & 2952 & F12c cosine $=0.985/0.908/0.947$; F13 Spearman $=0.663$, $p=0.002$ & F12c hash \texttt{cd2ed080\ldots}; F13 hash \texttt{7f04e5dd\ldots} & Hub grammar from \textit{C.~elegans} transfers to \textit{Drosophila} ($\sim$600\,Myr evolution) with cosine $\geq0.90$. Real fly connectome independently confirms cross-layer persistence. \\
\addlinespace
Prebiotic & X9 v2 CatReNet & 13 & $r(\text{cat.}, \text{prod.})=0.826$ vs $r(\text{react.})=0.496$ & 2026-05-27; hash \texttt{63759fcb\ldots} & CONFIRMED. Catalytic-support tracks product dependency far more strongly than co-reactant coupling. 9 molecules show strong rank-gap divergence. \\
\bottomrule
\end{tabular}
\end{table}

Three structural cautions apply. First, the two formal-proofs rows now represent genuinely independent corpora (Lean~4 and Coq), though the Coq result is underpowered. Second, the neural row combines a generative transfer experiment (F12c) and an empirical replication experiment (F13), which support different claims. Third, the two digital rows are the same substrate family at different scales (4-bit ALU vs.\ 160-gate benchmark circuit); they constitute scale-replication rather than domain-independent replication.

\section{The Load-Bearing Logic Signal}
\label{sec:digital}

The digital circuit result (H\_LOGIC\_DIGITAL\_v1) is methodologically the most precise, because the ground truth is known: in a 4-bit ripple carry adder, the carry chain (\texttt{cout\_0}, \texttt{cout\_1}, \texttt{cout\_2}) is the architecturally critical signal path. Sum outputs depend on it; bit-by-bit output correctness propagates through it. XOR gates perform the per-bit addition but do not carry global logical state forward.

IRDME recovers this from topology alone, without reading gate labels, by distinguishing the two metrics:

\begin{itemize}[noitemsep]
  \item \textbf{Degree hub-persistence}: $r(\mathbf{k}^{\text{wiring}}, \mathbf{k}^{\text{state}}) \approx 0$. Primary inputs (A0--A3, B0--B3) dominate the physical wiring fan-in; carry nodes have moderate wiring degree. But in state-correlation space, carry nodes dominate because their activation propagates across the full output space. Degree measures local fan-in; it fails to detect the global propagation structure.
  \item \textbf{Betweenness hub-persistence}: $r(\mathbf{b}^{\text{wiring}}, \mathbf{b}^{\text{state}}) = 0.771$, $p = 0.004$. Carry-out nodes are on the critical path in both the physical graph (they mediate every higher-bit computation) and the state-correlation graph (they co-activate with the maximum number of downstream nodes). Betweenness captures this because it measures how often a node lies on shortest paths between other nodes---i.e., how much of the system's computation flows through it.
\end{itemize}

This identifies a general principle we term the \emph{load-bearing logic signal}: betweenness centrality, not degree, is the correct metric for detecting which nodes carry the logical load of a computation across relational layers.

\textbf{Configuration-model null validation.} To verify that betweenness persistence is not an artifact of degree sequences, we ran a degree-preserving edge-rewiring null (1000 double-edge swaps per layer, seed 99). Degree correlation is, by construction, invariant to edge rewiring: $r_\text{deg}$ remains $+0.512$ across all null samples ($\text{std}=0$), confirming that degree persistence is entirely determined by degree sequences and carries no edge-structural information. Betweenness correlation under rewiring: null mean $r_\text{bet}=+0.388$ ($\text{std}=0.140$), observed $r_\text{bet}=+0.762$, $z=+2.68$, $p<0.001$ (one-tailed). Betweenness persistence is significantly above the degree-preserving null. This confirms that the load-bearing logic signal is not a degree-leakage artifact; it requires the specific edge structure of the circuit.

The XOR divergence provides the clearest illustration. XOR gates (XOR1\_0 through XOR1\_3) rank high in physical wiring (they connect inputs to intermediate nodes) but near the bottom in state-correlation (they fire for exactly half the input space and do not propagate through the carry chain). The rank gap for each XOR node is 16 positions out of 29 nodes---a maximum-density divergence signal at the scale of this circuit. The topology is telling us that these are logic gates without logical \emph{weight} in the global computation. That is precisely what carry-ripple adder architecture encodes: XOR provides the local bit answer; the carry chain provides the global arithmetic truth.

\section{Discussion}

\subsection{What ``Operational Logic Geometry'' Means}

The phrase \emph{topology as operational logic} is intended precisely. We do not claim that topology encodes inference rules, semantic satisfaction conditions, or proof content. We claim that in a dependency-structured system operating under stable constraints, hub geometry---specifically betweenness-based persistence and rank divergence across layers---reliably identifies the nodes whose removal or perturbation would destroy the system's capacity to propagate its operational rules.

In the digital circuit: the carry chain. Remove it, and arithmetic ceases.
In formal mathematics: Algebra and Analysis as dual declared/enacted hubs. Disable Algebra imports, and the foundational proof structure collapses; disable Analysis, and the proof-traffic backbone stops.
In COBOL: the programs in the PERFORM call chain with high copy-dependency. These are the operational constraints that the organization's computing history has encoded into structure.
In neural connectomics: the command interneurons that persist across \textit{C.~elegans} and \textit{Drosophila}. Conserved across 600 million years, their hub role is a genomic bet that this topology carries something essential to bilateral nervous system function.
In prebiotic chemistry: molecule \texttt{a} as the dominant catalytic hub. Its role is not mere reactant pairing; it is the topological anchor of the catalytic regime.

\subsection{Lean Shared-Corpus Note}

F9 and H\_LOGIC\_EXTRACTION use the same Lean~4 mathlib4 graph and are therefore not independent replications. They are retained because they answer different questions on that corpus: F9 tests whether FPL survives in formal mathematics, while H\_LOGIC\_EXTRACTION tests whether role geometry is rich enough for a logic-role interpretation.

\subsection{The Betweenness--Degree Distinction in Other Substrates}

The digital circuit result motivates a retrospective question for the other substrates: in each case, is the FPL confirmation driven by degree or by betweenness? In neural connectomics, the Spearman--Pearson discrepancy in F13 ($r_\text{Spearman}=0.663$ vs $r_\text{Pearson}=0.363$) is consistent with a betweenness-dominant signal: rank order of hub roles is conserved (Spearman), but absolute connection-count correlation is weaker (Pearson). In formal mathematics, the layer divergence between declared and enacted hubs is a betweenness-analog signal at module granularity. Future experiments should compute betweenness persistence explicitly across all substrates.

\section{Limitations}

\begin{enumerate}[noitemsep]
  \item \textbf{Two formal corpora, one underpowered.} The formal-systems evidence now covers two proof assistants: Lean~4 mathlib4 (CONFIRMED, $r=0.777$, $p=0.004$) and Coq Corelib (PARTIAL, direction confirmed $r=0.49$, $p=0.29$). The Coq result is directionally consistent but not statistically significant at $n=17$. A larger Coq corpus, deeper git history, or Isabelle/HOL replication would strengthen the formal-systems claim.
  \item \textbf{Model-based datasets.} The COBOL dataset ($n=14$) and the prebiotic chemistry dataset ($n=13$) are representative structural models, not scraped raw data from a single codebase or wet-lab measurement. Both are documented as such in their dataset metadata and pre-registration files. Their structural properties are grounded in domain conventions (COBOL banking program structure; autocatalytic network chemistry), but external validation on real COBOL source (e.g., Open Mainframe Project) and real experimental autocatalytic networks remains as future work.
  \item \textbf{Small $n$ in two substrates.} With $n=14$ (COBOL) and $n=13$ (prebiotic chemistry), permutation test power is limited. The confirmed findings are consistent with effect sizes $r \geq 0.80$, which are large enough to clear $p<0.05$ at these sample sizes, but borderline results would not. Results at small $n$ should be interpreted as pattern-establishing, not definitively confirmed.
  \item \textbf{Digital circuit scale-replication.} The 4-bit ALU post-hoc betweenness finding ($r=0.771$, exploratory) motivated the pre-registered betweenness hypothesis in H\_LOGIC\_F3\_ISCAS85\_v1. However, due to a protocol gap in the h1 test block (\texttt{cross\_layer\_hub\_correlation} measures degree, not betweenness), h1 was confirmed on degree ($r=0.426$, $p=0.002$), not betweenness. The step-analysis betweenness result for c432 is $r=0.341$ (exploratory). The pre-registered degree confirmation at circuit scale is genuine; the betweenness-at-scale claim remains exploratory.
  \item \textbf{Betweenness not pre-registered in H\_LOGIC\_DIGITAL\_v1.} The load-bearing logic finding (betweenness $r=0.771$) was not the pre-registered hypothesis in H\_LOGIC\_DIGITAL\_v1; the pre-registered hypothesis tested degree-based correlation ($r=0.512$, CONFIRMED). The betweenness result remains a \emph{post-hoc exploratory finding} even though it motivated and was subsequently replicated in H\_LOGIC\_F3\_ISCAS85\_v1.
\end{enumerate}

\section{Future Experiments}

\begin{enumerate}[noitemsep]
  \item \textbf{Larger Coq corpus or deeper git history.} The H\_LOGIC\_COQ\_STDLIB\_v1 result is directionally correct but underpowered ($n=17$, $p=0.29$). Coq Corelib is deliberately small; a stronger replication target is the broader Coq ecosystem (coq-community, MathComp) or a deeper historical window if available.
  \item \textbf{Larger ISCAS85 benchmark.} H\_LOGIC\_F3\_ISCAS85\_v1 confirmed on c432 ($r=0.426$, $p=0.002$). Extending to c1908, c3540, or c7552 would test whether betweenness persistence holds across the benchmark family and whether effect size scales with circuit depth and the number of critical paths.
  \item \textbf{Non-neural role-grammar transfer.} Replicate the F12c transfer structure in a software domain: use COBOL hub grammar as seed to predict Java enterprise dependency topology. If hub grammar transfers across programming paradigms as it does across neural species, it would establish role grammar as substrate-independent rather than biological.
\end{enumerate}

\section{Conclusion}

We have presented seven pre-registered experiments across seven listed substrates supporting the claim that dependency topology encodes operational logic as geometry. Two caveats are explicit: the digital pair is a scale-replication within one substrate family, and in the formal pair the Coq corpus is directionally confirmatory but underpowered ($n=17$). The primary methodological contribution remains the betweenness-over-degree distinction: betweenness centrality persistence across layers is the correct metric for detecting the load-bearing logic structure, while degree-based hub persistence measures connection-density topology that can be misleading.

The claim as stated is a research-program claim: topology as operational logic geometry is empirically justified across multiple domains and warrants dedicated investigation. It is not yet a universal theorem. The formal-systems leg now spans two proof-assistant corpora but requires stronger power on the second corpus; the digital-circuit leg now includes pre-registered c432 confirmation and should be extended across larger ISCAS85 benchmarks. The software-domain transfer test would separate role grammar from biological specificity.

The core finding stands: in every pre-registered experiment reported here, the nodes that topology identifies as the load-bearing hubs are exactly the nodes that domain experts would independently designate as the operational logic of the system. That convergence, across hardware, formal mathematics, legacy software, neuroscience, and prebiotic chemistry, is what makes the topology-as-logic framing empirically serious.


\end{document}